\documentclass[prd,nofootinbib,showpacs,preprintnumbers,amsmath,amssymb]{revtex4}
\usepackage{graphicx}% Include figure files
\usepackage{dcolumn}% Align table columns on decimal point
\usepackage{bm}% bold math

\def\bear{\begin{eqnarray}}
\def\ear{\end{eqnarray}}

\begin{document}

\title{Landau Levels of Scalar QED in Time-Dependent Magnetic Fields}

\author{Sang Pyo Kim}\email{sangkim@kunsan.ac.kr}
\affiliation{Department of Physics, Kunsan National University, Kunsan 573-701, Korea}

\medskip

\date{\today}

\begin{abstract}
The Landau levels of scalar QED undergo continuous transitions under a homogeneous, time-dependent magnetic field. We analytically formulate the Klein-Gordon equation for a charged spinless scalar as a Cauchy initial value problem in the two-component first order formalism and then put forth a measure that classifies the quantum motions into the adiabatic change, the nonadiabatic change, and the sudden change. We find the exact quantum motion and calculate the pair-production rate when the magnetic field suddenly changes as a step function.
\end{abstract}
\pacs{11.15.Tk, 12.20.Ds, 12.20.-m, 13.40.-f}

\maketitle

\section{Introduction} \label{sec1}

The interaction of charged particles with background electromagnetic fields has been intensively studied since the seminal works by Heisenberg and Euler, and Weisskopf on the effective action in a constant electromagnetic field in spinor QED \cite{heisenberg-euler36} and scalar QED \cite{weisskopf36}. Schwinger's proper-time formalism has laid a corner stone for probing vacuum polarization in the constant electromagnetic field \cite{schwinger51}. The quantum states of a charged particle are an essential ingredient in understanding the strong electromagnetic interaction and the vacuum polarization, in particular, in strongly magnetized neutron stars \cite{harding-lai06} and in extremely high-intensity lasers \cite{DMHK12}.

In a constant magnetic field charged spin-1/2 fermions or spinless scalars have Landau levels \cite{tsai-yildiz71,goldman-tsai71,bagrov-gitman,KLL02} and the energy spectrum leads to the QED effective action via the zeta-function regularization \cite{dunne04}. However, the quantum motion becomes non-trivial when the background electromagnetic field is localized in space or depends on time. Recently the effective actions in spinor or scalar QED have been found for the Sauter-type electric field localized in time or space, in which the quantum states of charged spin-1/2 fermion or spinless scalar are used in the in-out formalism \cite{KLY08,KLY10}, and the effective action has been found for the Sauter-type magnetic field localized in space in Ref. \cite{kim11}.

The resolvent method has also been used to compute the effective action in the Sauter-type electric field \cite{dunne-hall98} and the magnetic field \cite{dunne-hall98b}. The fermionic determinants have been investigated in static inhomogeneous magnetic fields \cite{fry95,fry96}. For numerical purposes, the worldline formalism has also been applied to scalar QED in a static magnetic field of step function \cite{gies-langfeld} and the real space split operator has been introduced for scalar QED in arbitrary electromagnetic potential \cite{RBK09}.
Hence analytical studies of quantum states of charged particles beyond the Sauter-type electric or magnetic field or in time-dependent magnetic fields will be not only of theoretical interest but also of practical applicability.

In this paper we study the quantum states of charged spinless scalars in homogeneous, time-dependent magnetic fields. For that purpose we observe that the Klein-Gordon (KG) equation in a time-dependent magnetic field and the Wheeler-DeWitt (WDW) equation for the Friedman-Robertson-Walker universe with a minimal massive scalar field have the same mathematical structure: a relativistic wave equation with a transverse motion of time-dependent harmonic oscillator. The Cauchy initial value problem for the WDW equation with a general scalar field has been studied in the two-component first order formalism \cite{kim91,kim-page92,kim92} and in the third quantization \cite{kim13}. Feshbach and Villars have introduced another two-component first order formalism \cite{feshbach-villars}, which was applied to the KG equation in Refs. \cite{mostafazadeh06a,mostafazadeh06b}.

The instantaneous Landau levels in time-dependent magnetic field do not decouple the KG equation and undergo continuous transitions among themselves during the quantum evolution. We formulate the Cauchy initial value problem, which expresses the KG equation as the two-component first order equation and incorporates the rate of change of Landau levels in addition to the instantaneous energy spectrum. The ratio of the rate of change of each Landau level to the corresponding dynamical phase during any time interval may provide a measure that characterizes and classifies the quantum motions of charged particles into (i) the adiabatic change, (ii) the sudden change, and (iii) the nonadiabatic change.

In the case of an adiabatic change of magnetic field the Landau levels change so slowly that the charged scalar remains in the same time-dependent Landau level and the adiabatic theorem holds, as expected by physical intuition. By contrast, in the case of a sudden change the Landau levels change so rapidly that the charged scalar cannot follow the time-dependent Landau level and is thus frozen to the initial Landau level. The magnetic field that suddenly changes as a step function provides an exactly solvable model, in which the dynamical phase is extremely small compared to a finite change of Landau levels during the infinitesimal interval and thus the measure becomes arbitrary large. In the last case of a nonadibatic change the charged scalar makes continuously transitions among time-dependent Landau levels, keeping the parity of the initial Landau level.

The organization of the paper is as follows. In Sec. \ref{sec2} we introduce the two-component first order formalism for the KG equation in time-dependent magnetic fields. Decomposing the field by the instantaneous Landau levels, we formulate the Cauchy initial value problem that evolves an initial Landau level. The two-component propagator is entirely determined by the rate of change of the instantaneous Landau levels and the energy spectrum. In Sec. \ref{sec3} we further propose a dimensionless measure which depends on the relative ratio of the rate of change of Landau levels to the dynamical phase over the time interval under study and which classifies the quantum motions into the adiabatic change, the sudden change, and the nonadiabatic change. In Sec. \ref{sec4} we find the quantum states of charged scalars and calculate the pair-production rate from the Dirac sea when the magnetic field abruptly changes as a step-function. In Sec. \ref{sec5} we discuss the physical implications of the quantum states in time-dependent magnetic fields.

\section{Quantum Motion of Scalar} \label{sec2}

We study the quantum states of a charged spinless scalar in the vector potential
\begin{eqnarray}
{\bf A} (t, {\bf x}) = \frac{1}{2} {\bf B} (t) \times {\bf x}, \label{vec gaug}
\end{eqnarray}
which leads to both the homogeneous, time-dependent magnetic field ${\bf B} (t)$ and the electric field ${\bf E} (t, {\bf x}) = - \partial {\bf A}/\partial t$.
The magnetic field is assumed to be along the $z$-direction and to have positive magnitude $|{\bf B}(t)|$ to guarantee the Landau levels.
The KG equation for a charge $q$ with mass $m$ in the vector potential (\ref{vec gaug}), after
decomposing the longitudinal Fourier-mode, takes the transverse motion (in units of $\hbar = c =1$)
\begin{eqnarray}
\Biggl[\frac{d^2}{d t^2} + \hat{\rm p}_{\perp}^2 + \Bigl(\frac{qB(t)}{2} \Bigr)^2 \hat{\bf x}^2_{\perp} - q B (t) \hat{L}_z  + k_z^2 + m^2 \Biggr] \Psi_{\perp} (t, {\bf x}_{\perp}) = 0, \label{kg eq}
\end{eqnarray}
where $\hat{\bf p}_{\perp} = - i \nabla$ and $\hat{L}_z = \hat{\rm x}_{\perp} \times \hat{\rm p}_{\perp}$.

Note that the transverse Hamiltonian
\begin{eqnarray}
\hat{H}_{\perp} (t) = \hat{\bf p}_{\perp}^2 + \Bigl(\frac{qB (t)}{2} \Bigr)^2 \hat{\bf x}^2_{\perp} - q B (t) \hat{L}_z \label{tran ham}
\end{eqnarray}
describes a two-dimensional time-dependent oscillator coupled to the angular momentum $\hat{L}_z$.
In the oscillator representation
\begin{eqnarray}
\hat{a}_x (t) = \frac{\sqrt{qB(t)}}{2} \hat{x} + \frac{2i}{\sqrt{qB (t)}} \hat{p}_x, \quad \hat{a}^{\dagger}_x (t) = {\rm H. C. },  \nonumber\\
\hat{a}_y (t) = \frac{\sqrt{q B (t)}}{2} \hat{y} + \frac{2i}{\sqrt{qB (t)}} \hat{p}_y, \quad \hat{a}^{\dagger}_y (t) = {\rm H. C. }, \label{osc rep}
\end{eqnarray}
the transverse Hamiltonian (\ref{tran ham}) is given by
\begin{eqnarray}
\hat{H}_{\perp} (t) = qB (t) \bigl[ \hat{a}^{\dagger}_x (t) \hat{a}_x (t) + \hat{a}^{\dagger}_y (t) \hat{a}_y (t) + 1 \bigr] + i qB (t)
\bigl[ \hat{a}^{\dagger}_x (t) \hat{a}_y (t) - \hat{a}_x (t) \hat{a}^{\dagger}_y (t) \bigr]. \label{a ham}
\end{eqnarray}
Further, in  the new basis \cite{wybourne}
\begin{eqnarray}
\hat{c}_{\pm } (t) = \frac{1}{\sqrt{2}} \bigl( \hat{a}_x (t) \mp i \hat{a}_y (t) \bigr) \label{diag basis}
\end{eqnarray}
with the equal-time commutators
\begin{eqnarray}
[\hat{c}_{\pm} (t), \hat{c}^{\dagger}_{\pm} (t) ] = 1, \quad [\hat{c}_{\pm} (t), \hat{c}_{\mp} (t)] = [\hat{c}_{\pm} (t), \hat{c}^{\dagger}_{\mp} (t)] = 0,
\end{eqnarray}
Eq. (\ref{a ham}) can be written in the diagonal form
\begin{eqnarray}
\hat{H}_{\perp} (t) =  q B(t) \bigl[ \hat{c}^{\dagger}_{-} (t) \hat{c}_{-} (t) + \hat{c}_{-} (t) \hat{c}^{\dagger}_{-} (t) \bigr]. \label{diag ham}
\end{eqnarray}
Hence the Landau levels for Eq. (\ref{diag ham}) are the number states of $\hat{c}_{-}^{\dagger}(t) \hat{c}_{-}(t)$:
\begin{eqnarray}
\hat{c}_{-} (t) \vert 0, t \rangle = 0, \quad
\vert n, t \rangle = \frac{(\hat{c}^{\dagger}_{-} (t))^n}{\sqrt{n!}} \vert 0, t \rangle. \label{num st}
\end{eqnarray}
However, note that the Landau levels (\ref{num st}) do not separate Eq. (\ref{kg eq}) into a diagonal one since they explicitly depend on time and make continuous transitions among themselves.

The Landau levels, arranged into a column vector in increasing quantum numbers,
\begin{eqnarray}
\vec{\Phi} (t) = \begin{pmatrix}
  \vert 0, t \rangle \\
\vert 1, t \rangle \\
\vdots
 \end{pmatrix}, \label{num rep}
\end{eqnarray}
change as
\begin{eqnarray}
\frac{d}{d t} \vec{\Phi} (t) = \Omega (t) \vec{\Phi} (t). \label{tran rat}
\end{eqnarray}
Here the rate of change of the basis has the oscillator representation
\begin{eqnarray}
\Omega (t) = \frac{\dot{B} (t)}{4 B (t)} \bigl(\hat{c}^2_{-} (t) - \hat{c}^{\dagger 2}_{-} (t) \bigr), \label{Omega}
\end{eqnarray}
and the matrix representation
\begin{eqnarray}
\langle m, t \vert \Omega (t) \vert n, t \rangle = \frac{\dot{B} (t)}{4 B (t)} \bigl(\sqrt{n(n-1)} \delta_{m, n-2} - \sqrt{(n+1)(n+2)} \delta_{m, n+2} \bigr). \label{Omega-mat}
\end{eqnarray}
Thus the Landau levels unitarily transform as
\begin{eqnarray}
 \vec{\Phi} (t) = {\bf S} (t, t_0) \vec{\Phi} (t_0), \label{tran mat}
\end{eqnarray}
where the transition matrix in the time-ordered integral yields a one-mode squeeze operator \cite{stoler70}
\begin{eqnarray}
{\bf S}(t, t_0) = {\cal T} \exp \Bigl[ \int_{t_0}^{t} \Omega (t') dt' \Bigr] = \exp \Bigl[\frac{1}{4} \ln \Bigl(\frac{B(t)}{B(t_0)} \Bigr) (\hat{c}^2_{-} (t_0) - \hat{c}^{\dagger 2}_{-} (t_0) ) \Bigr]. \label{tran mat}
\end{eqnarray}
Following  Refs. \cite{kim91,kim-page92,kim92}, we may expand the field as
\begin{eqnarray}
\Psi_{\perp} (t, {\bf x}_{\perp})  = \vec{\Phi}^{T} (t, {\bf x}_{\perp}) {\bf S} (t, t_0) \vec{\Psi}_{\perp} (t) \label{field ex}
\end{eqnarray}
with $\vec{\Phi}^{T} (t, {\bf x}_{\perp})$ denoting the transpose of the coordinate representation of the Landau levels (\ref{num rep}).
We then write Eq. (\ref{kg eq}) in the two-component first order formalism
\begin{eqnarray}
\frac{d}{d t}
\begin{pmatrix}
\Psi_{ \perp} (t) \\
\frac{d \Psi_{ \perp} (t) }{d t} \end{pmatrix}
= \begin{pmatrix}
  0 & I \\
 - {\bf S}^{-1}(t, t_0) {\bf \omega}^2 (t) {\bf S} (t,t_0)  & 0 \end{pmatrix}
 \begin{pmatrix}
\Psi_{ \perp} (t) \\
\frac{d \Psi_{ \perp} (t)}{d t} \end{pmatrix}, \label{kim eq}
\end{eqnarray}
where $I$ is the identity matrix and
\begin{eqnarray}
{\bf \omega}^2 (t) = qB (t) (2\hat{c}^{\dagger} (t) \hat{c} (t) + 1) + m^2 + k_z^2 \label{freq}
\end{eqnarray}
is the diagonal matrix of the Landau energy. Remarkably, the Cauchy data are given by \cite{kim91,kim-page92,kim92}
\begin{eqnarray}
\begin{pmatrix}
\Psi_{\perp} (t, {\bf x}_{\perp})  \\
\frac{\partial \Psi_{\perp} (t, {\bf x}_{\perp})}{\partial t} \end{pmatrix}
= \begin{pmatrix}
 \vec{\Phi}^T (t, {\bf x}_{\perp})  & 0 \\
 0 & \vec{\Phi}^T (t, {\bf x}_{\perp}) \end{pmatrix} {\cal U} (t, t_0) \begin{pmatrix}
\vec{\Psi}_{\perp} (t_0)\\
\frac{d \vec{\Psi}_{\perp} (t_0)}{d t}  \end{pmatrix}, \label{cauchy}
\end{eqnarray}
where ${\bf S}^{-1}$ from Eq. (\ref{kim eq}) cancels ${\bf S}$ in Eq. (\ref{field ex}) and the two-component propagator can be given by the time-ordered integral
\begin{eqnarray}
{\cal U} (t, t_0) = {\cal T}
 \exp \Biggl[ \int_{t_0}^{t} \begin{pmatrix}
  \Omega (t') & I \\
 - \omega^2 (t') & \Omega (t') \end{pmatrix} d t' \Biggr]. \label{two prop}
\end{eqnarray}
The Cauchy data (\ref{cauchy}) guarantees the inner product between the positive and negative frequency solutions for the KG equation
\begin{eqnarray}
i \int d^3{\bf x} \Bigl(\Psi^{(-)} (t, {\bf x}) \frac{\partial}{\partial t} \Psi^{(+)} (t, {\bf x}) - \Psi^{(+)} (t, {\bf x}) \frac{\partial}{\partial t} \Psi^{(-)} (t, {\bf x}) \Bigr) = I. \label{in prod}
\end{eqnarray}

\section{Classification of Quantum Motions} \label{sec3}

As the evolution (\ref{two prop}) is carried by ${\bf \omega}^2$ and $\Omega$, we may put forth a dimensionless measure that characterizes the quantum motion of the $n$th Landau level during any time interval $(t_i, t_f)$ as
\begin{eqnarray}
{\cal R}_n = \frac{\frac{n}{4} \Bigl| \ln \Bigl(\frac{B(t_f)}{B(t_i)} \Bigr) \Bigr|}{\int_{t_i}^{t_f} {\bf \omega} (t', n) dt'}. \label{R}
\end{eqnarray}
Note further that the Landau energy (\ref{freq}) in the $n$th Landau level is dominated by the mass $m$ when the magnetic field has an under-critical strength, $nB(t)/m^2 \ll 1$, whereas it is dominated by $nB(t)$ when the field has an over-critical strength, $nB(t)/m^2 \gg 1$. Since the lower bound for the dynamical phase is $m \Delta t$ for $\Delta t = t_f - t_i$, the upper bound for the measure ${\cal R}_n$ is
\begin{eqnarray}
{\cal R}_n \leq \frac{\frac{n}{4} \Bigl| \ln \Bigl(\frac{B(t_f)}{B(t_i)} \Bigr) \Bigr|}{m \Delta t}.
\end{eqnarray}
Hence we may classify the quantum motions into three categories: (i) the adiabatic change when ${\cal R}_n \ll 1$, (ii) the sudden change when ${\cal R}_n \gg 1$, and (iii) the nonadiabatic change, otherwise. An interesting model is provided by a modulated magnetic field in a constant background
\begin{eqnarray}
B(t) = B_0 + B_1 \cos \Bigl( \frac{t}{T} \Bigr), \quad (B_0 > B_1).
\end{eqnarray}
The background field should be larger than the modulated field in order to exclude the tachyonic states for the over-critical strength, but not necessarily, otherwise. For under-critical strengths $B_0, B_1 \ll m$, the measure is given by
\begin{eqnarray}
{\cal R}_n \approx \frac{\frac{n}{4} \Bigl| \ln \Bigl(\frac{B_0 + B_1}{B_0 - B_1} \Bigr) \Bigr|}{m \Delta T}.
\end{eqnarray}
The measure can be made large by choosing a very small $T$ and  $\Delta B = B_0 - B_1$ in the Compton scale for $m$. For instance, ${\cal R}_n = 1$ requires $\ln (B_0/\Delta B)/T = 10^{20}$ for electrons and positrons.

In the first case (i) of the adiabatic change, $\Omega$ can be neglected and the two-component propagator is approximately given by
\begin{eqnarray}
{\cal U} (t, t_0) \approx {\cal P} (t, t_0), \label{i}
\end{eqnarray}
where
\begin{eqnarray}
{\cal P} (t, t_0) = {\cal T}
 \exp \Biggl[ \int_{t_0}^{t} \begin{pmatrix}
  0 & I \\
 - {\bf \omega}^2 (t') & 0 \end{pmatrix} d t' \Biggr]. \label{P diag}
\end{eqnarray}
Note that Eq. (\ref{P diag}) can be written as
\begin{eqnarray}
{\cal P} (t, t_0) = {\cal P} (t) {\cal P}^{-1} (t_0), \quad {\cal P} (t) = \begin{pmatrix}
P_1 (t) & P_2 (t) \\  \dot{P}_1 (t) & \dot{P}_2 (t) \end{pmatrix} \label{p mat}
\end{eqnarray}
where $P_1(t)$ and $P_2 (t)$ are two independent solutions to the diagonal matrix equation
\begin{eqnarray}
\frac{d^2 P(t)}{dt^2} + {\bf \omega}^2 (t) P(t) = 0. \label{diag eq}
\end{eqnarray}
The charged scalar remains in the same Landau level which adiabatically changes when the magnetic field slowly changes, and thus the adiabatic theorem holds.

In the second case (ii) of the sudden change, $\Omega$ dominates over $\omega$ and $I$, so the two-component propagator is approximately given by
\begin{eqnarray}
{\cal U} (t, t_0) \approx
 \begin{pmatrix}
 {\bf S} (t, t_0) & 0 \\
 0 & {\bf S} (t, t_0) \end{pmatrix}. \label{ii}
\end{eqnarray}
The one-mode squeeze operator ${\bf S}$ in Eq. (\ref{ii}) cancels another ${\bf S}^{\dagger}$ from $\vec{\Phi}^T (t, {\bf x}_{\perp})= \vec{\Phi}^T (t_0, {\bf x}_{\perp}) {\bf S}^{\dagger}$ in Eq. (\ref{tran mat}), so the Cauchy data are approximately given by
\begin{eqnarray}
\begin{pmatrix}
\Psi_{\perp} (t, {\bf x}_{\perp})  \\
\frac{\partial \Psi_{\perp} (t, {\bf x}_{\perp})}{\partial t} \end{pmatrix}
\approx \begin{pmatrix}
\vec{\Phi}^T (t_0, {\bf x}_{\perp}) \cdot \vec{\Psi}_{\perp} (t_0)\\
\vec{\Phi}^T (t_0, {\bf x}_{\perp}) \cdot \frac{d \vec{\Psi}_{\perp} (t_0)}{d t}  \end{pmatrix}. \label{iii-st}
\end{eqnarray}
The charged scalar does not follow the time-dependent Landau level and is frozen to the initial one when the magnetic field suddenly changes such that ${\cal R}_n \gg 1$. In Sec. \ref{sec4} we shall consider the most typical model of this category, in which the magnetic field suffers a step-function change and ${\cal R}_n = \infty$.

In the third case (iii) of the nonadiabatic change, in which $\Omega$ is comparable to ${\bf \omega}$, using the similarity formula \cite{dollard-friedman77,dollard-friedman79,kim92}, we may write the propagator in terms of the two-component propagator (\ref{p mat})
\begin{eqnarray}
{\cal U} (t, t_0) = {\cal P} (t) {\cal T}
 \exp \Biggl[ \int_{t_0}^{t} {\cal P}^{-1} (t') \begin{pmatrix}
  \Omega (t') & 0 \\
 0 & \Omega (t') \end{pmatrix} {\cal P}(t') d t' \Biggr] {\cal P}^{-1} (t_0). \label{prop}
\end{eqnarray}
The equivalence between Eqs. (\ref{two prop}) and (\ref{prop}) can be shown by taking a derivative with respect to time. Now the charged scalar makes continuous transitions among Landau levels due to the transition matrix from the time-ordered integral in Eq. (\ref{prop}), for instance, the first two terms
\begin{eqnarray}
{\cal U} (t, t_0) = {\cal P} (t) \Biggl[ I +
 \int_{t_0}^{t} {\cal P}^{-1} (t') \begin{pmatrix}
  \Omega (t') & 0 \\
 0 & \Omega (t') \end{pmatrix} {\cal P}(t') d t' + \cdots \Biggr]{\cal P}^{-1} (t_0).
\end{eqnarray}
The first term is the adiabatic evolution and the second term comes from continuous transitions of Landau levels. The nonperturbative form may be found using the Magnus expansion \cite{BCOR09,kim92}, which is beyond the scope of this paper and will be separately treated elsewhere.

\section{Sudden Change Model} \label{sec4}

As a solvable model, we consider a sudden change in which the magnetic field jumps from $B_0$ to $B_1$ with a step function
\begin{eqnarray}
B(t) = (B_1 - B_0) \theta (t) + B_0. \label{sud B}
\end{eqnarray}
The magnetic field (\ref{sud B}) is a mathematical  model in that the induced electric field has a delta function profile.
We denote $\hat{c}_{\rm in}$ when $t < 0$ and $\hat{c}_{\rm out}$ when $t > 0$ for the basis (\ref{diag basis}) for the Landau levels.
Since the rate of change $\Omega$ in Eq. (\ref{Omega}) is proportional to $\delta (t)$, the transition matrix (\ref{tran mat}) is the one-mode squeeze operator
\begin{eqnarray}
{\bf S} = \exp \Bigl[ \frac{1}{4} \ln \Bigl(\frac{B_1}{B_0} \Bigr) (\hat{c}^{2}_{\rm in} - \hat{c}^{\dagger 2}_{\rm in}) \Bigr]. \label{sq op}
\end{eqnarray}
Alternatively, the Bogoliubov transformation
\begin{eqnarray}
\hat{c}_{\rm out} = \frac{1}{2} \Bigl(\sqrt{\frac{B_1}{B_0}} + \sqrt{\frac{B_0}{B_1}} \Bigr) \hat{c}_{\rm in}
 + \frac{1}{2} \Bigl(\sqrt{\frac{B_1}{B_0}} - \sqrt{\frac{B_0}{B_1}} \Bigr) \hat{c}^{\dagger}_{\rm in} \label{bog tran}
\end{eqnarray}
leads to the unitary transformation \cite{stoler70}
\begin{eqnarray}
\hat{c}_{\rm out} = {\bf S} \hat{c}_{\rm in} {\bf S}^{\dagger}, \quad
\hat{c}^{\dagger}_{\rm in} = {\bf S}^{\dagger} \hat{c}^{\dagger}_{\rm out} {\bf S}. \label{unit tran}
\end{eqnarray}
Hence each Landau level transforms into a squeezed one, $\vert n, {\rm out} \rangle = {\bf S} \vert n, {\rm in} \rangle$,
and similarly the column vector (\ref{num rep}) of Landau levels transforms as $\vec{\Phi}_{\rm out} = {\bf S} \vec{\Phi}_{\rm in}$ after the sudden change of magnetic field.

We decompose the two-component propagator (\ref{two prop}) into three parts
\begin{eqnarray}
{\cal U} (t, t_0) = {\cal P}(t, \epsilon; \hat{c}_{\rm out}, \hat{c}^{\dagger}_{\rm out}) {\cal T}
 \exp \Biggl[ \int_{-\epsilon}^{\epsilon} \begin{pmatrix}
  \Omega (t') & I \\  - {\bf \omega}^2 (t') & \Omega (t') \end{pmatrix} d t' \Biggr]  {\cal P} (-\epsilon, t_0; \hat{c}_{\rm in}, \hat{c}^{\dagger}_{\rm in}). \label{prop2}
\end{eqnarray}
The post-factor in Eq. (\ref{prop2}) is the propagator from the initial time $t_0$ to an infinitesimal time $-\epsilon$  in the basis of $\hat{c}_{\rm in}$ and $\hat{c}_{\rm in}^{\dagger}$, the mid-factor is the propagator from $- \epsilon$ to $\epsilon$, and the pre-factor is the propagator from $\epsilon$ to the final time $t$ in the basis of $\hat{c}_{\rm out}$ and $\hat{c}_{\rm out}^{\dagger}$.
In the limit of $\epsilon = 0$, evaluating the propagator by the transition matrix (\ref{sq op}) as
\begin{eqnarray}
{\cal T} \exp \Biggl[ \int_{-\epsilon}^{\epsilon} \begin{pmatrix}
  \Omega (t') & I \\  - {\bf \omega}^2 (t') & \Omega (t') \end{pmatrix} d t' \Biggr]  =
  \begin{pmatrix}
  {\bf S} & 0 \\  0 &  {\bf S} \end{pmatrix}, \label{S mat}
\end{eqnarray}
and using (\ref{unit tran}), we arrive at the Cauchy data
\begin{eqnarray}
\begin{pmatrix}
\Psi_{\perp} (t, {\bf x}_{\perp})  \\
\frac{\partial \Psi_{\perp} (t, {\bf x}_{\perp})}{\partial t} \end{pmatrix}
= \begin{pmatrix}
 \vec{\Phi}^T (t_0, {\bf x}_{\perp})  & 0 \\
 0 & \vec{\Phi}^T (t_0, {\bf x}_{\perp}) \end{pmatrix} {\cal P} (t, 0; \hat{c}_{\rm in}, \hat{c}^{\dagger}_{\rm in})
 {\cal P} (0, t_0; \hat{c}_{\rm in}, \hat{c}^{\dagger}_{\rm in}) \begin{pmatrix}
\vec{\Psi}_{\perp} (t_0)\\
\frac{d \vec{\Psi}_{\perp} (t_0)}{d t}  \end{pmatrix}. \label{sud-cauchy}
\end{eqnarray}
The Magnus expansion \cite{BCOR09} shows that Eq. (\ref{S mat}) is the correct limit since the leading correction of ${\cal O} (\epsilon^2)$  vanishes and the higher terms are order of ${\cal O} (\epsilon^3)$.

The two-component propagator
\begin{eqnarray}
{\cal P} (t) = \frac{1}{\sqrt{2 {\bf \omega}}} \begin{pmatrix} e^{- i {\bf \omega} t}  &  e^{i {\bf \omega} t} \\
 - i {\bf \omega} e^{- i {\bf \omega} t}  & i {\bf \omega}e^{ i {\bf \omega} t}\end{pmatrix},
\end{eqnarray}
consists of the positive and negative frequency solutions column-wise, respectively. Hence the charged scalar has the quantum state
\begin{eqnarray}
\Psi_{\perp} (t, {\bf x}_{\perp}) = \vec{\Phi}^T (t_0, {\bf x}_{\perp}) {\bf F} (t) \vec{\Psi}_{\perp} (t_0),
\end{eqnarray}
with the amplitude matrix
\begin{eqnarray}
{\bf F} (t) = \Bigl( \frac{{\bf \omega}_{\rm out} + {\bf \omega}_{\rm in}}{2
{\bf \omega}_{\rm out}} \Bigr) e^{- i {\bf \omega}_{\rm out}t + i {\bf \omega}_{\rm in} t_0 } + \Bigl( \frac{{\bf \omega}_{\rm out} - {\bf \omega}_{\rm in}}{2
{\bf \omega}_{\rm out}} \Bigr) e^{i {\bf \omega}_{\rm out}t +i {\bf \omega}_{\rm in} t_0 }, \label{amp}
\end{eqnarray}
where ${\bf \omega}_{\rm in}$ and ${\bf \omega}_{\rm out}$ are the diagonal matrix (\ref{freq}) with $B_0$ for $t< 0$ and $B_1$ for $t>0$, respectively.
As explained in Sec. \ref{sec3}, the charged scalar initially in the Landau level $\Phi_n (t_0, {\bf x}_{\perp})$ is frozen to that level with the time-dependent
amplitude ${\bf F}_n(t)$. The coefficients in Eq. (\ref{amp}) can also be obtained from the quantum scattering of a wave $e^{- i {\bf \omega}_{\rm in} t}$ into $e^{- i {\bf \omega}_{\rm out} t}$ and $e^{ i {\bf \omega}_{\rm out} t}$ by a potential step. Thus the temporal oscillation of $|{\bf F}_n (t)|^2$, as shown in Fig. 1, is a consequence of partial scattering of the positive frequency into the negative one after the sudden change of the energy.
 \begin{figure}[t]
%\subfloat[]
{\includegraphics[width=0.45\linewidth,height=0.25\textheight]{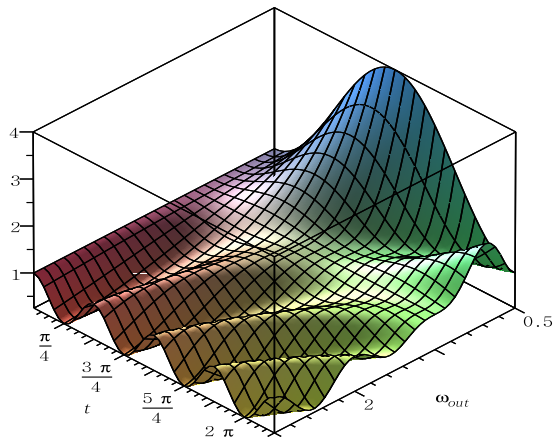} }\hfill

\caption{The magnitude $|F_n (t, {\bf \omega}_{\rm out})|^2$ is drawn in the range of $t = [0, \pi]$ and ${\bf \omega}_{\rm out} = [\frac{1}{2}, 2]$ in the scale of ${\bf \omega}_{\rm in} =1$.} \label{fig1}
\end{figure}
The probability interpretation of $|{\bf F}_n (t)|^2$ is not the correct prescription for the KG equation. Instead, the inner product (\ref{in prod}) with respect to the in-state and the out-state
\begin{eqnarray}
\vert {\rm in}, t \rangle =  \frac{e^{- i {\bf \omega}_{\rm in} t}}{\sqrt{2 {\bf \omega}_{\rm in}}}  \vec{\Phi} (t_0), \quad \vert {\rm out}, t \rangle =  \frac{e^{- i {\bf \omega}_{\rm out} t}}{\sqrt{2 {\bf \omega}_{\rm out}}} \vec{\Phi} (t_0),
\end{eqnarray}
leads to the Bogoliubov coefficients
\begin{eqnarray}
{\bf \alpha} = \frac{1}{2} \Bigl(\sqrt{\frac{{\bf \omega}_{\rm out}}{{\bf \omega}_{\rm in}}} + \sqrt{\frac{{\bf \omega}_{\rm in}}{{\bf \omega}_{\rm out}}} \Bigr), \quad
{\bf \beta} = \frac{1}{2} \Bigl(\sqrt{\frac{{\bf \omega}_{\rm out}}{{\bf \omega}_{\rm in}}} - \sqrt{\frac{{\bf \omega}_{\rm in}}{{\bf \omega}_{\rm out}}} \Bigr).
\end{eqnarray}
Considering the quantum motion of virtual charged scalar pairs in the Dirac sea, the pair-production rate is $|\beta (n)|^2$ for each Landau level. The induced electric field reinforces the argument of pair production from a time-dependent magnetic field. As far as pair production is concerned, time-dependent magnetic fields strongly contrast constant magnetic fields, in which the Dirac sea is stable at one loop and prohibits charged pairs from being emitted.
Though the direction of the magnetic field is fixed in this paper, it would be interesting to compare pair production from a rotating magnetic field in Ref. \cite{piazza-calucci02}.

\section{Conclusion} \label{sec5}

We have studied the quantum evolution of charged scalars in a homogeneous, time-dependent magnetic field. In contrast to a constant magnetic field, the Landau levels that instantaneously diagonalize the KG equation are a unitary transformation of initial ones via a one-mode squeeze operator when the magnetic field changes from a constant value. The two-component first order formalism has been employed to solve the Cauchy initial problem (\ref{cauchy}) in the basis of time-dependent Landau levels. We have introduced a dimensionless measure (\ref{R}) that classifies the quantum motions into three categories: (i) the adiabatic change, (ii) the sudden change, and (iii) the nonadiabatic change. The measure is the ratio of the rate of change of each Landau level to the corresponding dynamical phase for the time interval under study. When the magnetic field changes so slowly that the ratio is very small, the time-dependent Landau level adiabatically changes and the charged scalar remains in the same Landau level, while when the magnetic field changes so rapidly that the ratio is very large, the charged scalar cannot follow the rapidly changing Landau level and is frozen to the initial Landau level. On the other hand, when the ratio is order of unity and the rate of change of the Landau level is comparable to the corresponding energy, the charged scalar makes continuous transitions among time-dependent Landau levels, keeping the parity of the initial state, during the quantum evolution.

We have explicitly analyzed the quantum states for the charged scalar when the magnetic field changes from one constant value to another as a step function. The  two-component propagator (\ref{cauchy}) and (\ref{two prop}) during an infinitesimal interval for the change of the magnetic field reduces to the one-mode squeeze operator for the change of Landau levels, and the resulting quantum state remains the same initial Landau level. This implies that the scalar does not follow the instantaneously changing Landau level and is frozen in the initial Landau level. We have found the Bogoliubov coefficients when the out-state is the frozen Landau levels with the energy spectrum in the changed magnetic field. This implies that a time-dependent magnetic field may produce pairs of charged particles from the Dirac sea, which is supported from the presence of an induced electric field.

The use of the quantum states for vacuum polarization, pair production, and possible applications to astrophysics and extremely high-intensity lasers in general time-dependent magnetic fields are beyond the scope of this paper and will be addressed in a future publication.

\acknowledgments
The author thanks Hyun Kyu Lee for useful discussions on Landau levels, Berry's phases, and pair production in time-dependent magnetic fields. He also would like to thank Misao Sasaki and Takahiro Tanaka for the warm hospitality at the Yukawa Institute for Theoretical Physics of Kyoto University, where this paper was initiated during the Long-term Workshop YITP-T-12-03 on ``Gravity and Cosmology 2012,'' thank W-Y. Pauchy Hwang for the warm hospitality at the Center for Theoretical Sciences, Taipei, Taiwan, R.O.C. during the 4th APCosPA Winter School at National Taiwan University, where part of paper was done, and thank Don N. Page for the warm hospitality at University of Alberta, where this paper was revised.
This work was supported by Basic Science Research Program through the National Research Foundation of Korea (NRF) funded by the Ministry of Education (NRF-2012R1A1B3002852).


\begin{thebibliography}{99}


\bibitem{heisenberg-euler36} W.~Heisenberg and H.~Euler, ``Consequences of Dirac’s Theory of Positrons,'' Z.\ Phys.\ {\bf 98},  714 (1936).

\bibitem{weisskopf36} V.~Weisskopf, ``The electrodynamics of the vacuum based on the quantum theory of the electron,'' Kong.\ Dans.\ Vid.\ Selsk.\ Math-fys. Medd. {\bf XIV}  No. 6 (1936).

\bibitem{schwinger51} J.~Schwinger, ``On gauge invariance and vacuum polarization,'' Phys.\ Rev.\ {\bf 82}, 664 (1951).

\bibitem{harding-lai06} A.~K.~Harding and D.~Lai, ``Physics of strongly magnetized neutron stars,'' Rep.\ Prog.\ Phys.\ {\bf 69}, 2631 (2006).

\bibitem{DMHK12} A.~Di~Piazza, C.~M\"{u}ller†, K.~Z.~Hatsagortsyan and C.~H.~Keitel, ``Extremely high-intensity laser interactions with fundamental quantum systems,'' Rev.\ Mod.\ Phys.\ {\bf 84}, 1177 (2012).

\bibitem{tsai-yildiz71} W-Y.~Tsai and A.~Yildiz, ``Motion of Charged Particles in a Homogeneous Magnetic Field,'' Phys.\ Rev.\ D {\bf 4}, 3643 (1971).

\bibitem{goldman-tsai71} T.~Goldman and W-Y. Tsai,``Motion of Charged Particles in a Homogeneous Magnetic Field. II,'' Phys.\ Rev.\ D {\bf 4}, 3648 (1971).

\bibitem{bagrov-gitman} V.~G.~Bagrov and D.~M.~Gitman, {\it Exact Solutions of Relativistic Wave Equations} (Kluwer Academic, Dordrecht, 1990).

\bibitem{KLL02} S.~Kim, C.~Lee and K.~Lee, ``Quantum field dynamics in a uniform magnetic field: Description using fields in oblique phase space,'' Phys.\ Rev.\ D {\bf 65}, 045009 (2002).
\bibitem{dunne04} G.~V.~Dunne, ``Heisenberg-Euler Effective Lagrangians : Basics and Extensions,'' in {\it From Fields to Strings : Circumnavigating Theoretical
Physics}, edited by M.~Shifman, A.~Vainshtein and J.~Wheater (World Scientific, Singapore, 2004) [arXiv:hep-th/0406216].

\bibitem{KLY08} S.~P.~Kim, H.~K.~Lee and Y.~Yoon, ``Effective action of QED in electric field backgrounds,'' Phys.\ Rev.\ D {\bf 78}, 105013 (2008).

\bibitem{KLY10} S.~P.~Kim, H.~K.~Lee and Y.~Yoon, ``Effective action of QED in electric field backgrounds. II. Spatially localized fields,'' Phys.\ Rev.\ D {\bf 82}, 025015 (2010).

\bibitem{kim11} S.~P.~Kim, ``QED effective action in magnetic field backgrounds and electromagnetic duality,'' Phys.\ Rev.\ D {\bf 84}, 065004 (2011).

\bibitem{dunne-hall98} G.~Dunne and T.~Hall, ``QED effective action in time dependent electric backgrounds,'' Phys.\ Rev.\ D {\bf 58}, 105022 (1998).

\bibitem{dunne-hall98b} G.~Dunne and T.~M.~Hall, ``An exact ${\rm QED}_{3+1}$ effective action,'' Phys.\ Lett.\ B {\bf 419}, 322 (1998).

\bibitem{fry95} M.~P.~Fry, ``Fermion determinants in static, inhomogeneous magnetic fields,'' Phys.\ Rev.\ D {\bf 51}, 810 (1995).

\bibitem{fry96} M.~P.~Fry, ``QED in inhomogeneous magnetic fields,'' Phys.\ Rev.\ D {\bf 54}, 6444 (1996).

\bibitem{gies-langfeld} H.~Gies and K. Langfeld, ``Quantum diffusion of magnetic fields in a numerical worldline approach,'' Nucl.\ Phys.\ B {\bf 613}, 353 (2001).

\bibitem{RBK09} M.~Ruf, H.~Bauke and C.~H.~Keitel, ``A real space split operator method for the Klein–Gordon equation,'' J.\ Comp.\ Phys.\ {\bf 228}, 9092 (2009).

\bibitem{kim91} S.~P.~Kim, ``Quantum inflationary minisuperspace cosmological models,'' Ph.D. thesis, Pennsylvania State University, 1991.

\bibitem{kim-page92} S.~P.~Kim and D.~N.~Page, ``Wormhole spectrum of a quantum Friedmann-Robertson-Walker cosmology minimally coupled to a power-law scalar field and the cosmological constant,'' Phys.\ Rev.\ D {\bf 45}, 45, R3296 (1992).

\bibitem{kim92} S.~P.~Kim, ``Quantum mechanics of conformally and minimally coupled Friedmann-Roberston-Walker cosmology,'' Phys.\ Rev.\ D {\bf 46}, 3403 (1992).

\bibitem{kim13} S.~P.~Kim, ``Massive Scalar Field Quantum Cosmology,'' arXiv:1304.7439.

\bibitem{feshbach-villars} H.~Feshbach and F.~Villars, ``Elementary Relativistic Wave Mechanics of Spin 0 and Spin 1/2 Particles,'' Rev.\ Mod.\ Phys.\ {\bf 30}, 24 (1958)

\bibitem{mostafazadeh06a}  A.~Mostafazadeh and F.~Zamani, ``Quantum mechanics of Klein-Gordon fields I: Hilbert space, localized states, and chiral symmetry,'' Ann.\ Phys.\ {\bf 321}, 2183 (2006).

\bibitem{mostafazadeh06b}  A.~Mostafazadeh and F.~Zamani, ``Quantum mechanics of Klein-Gordon fields II: Relativistic cohrent states,'' Ann.\ Phys.\ {\bf 321}, 2210 (2006).

\bibitem{wybourne} G.~B.~Wybourne, {\it Classical Groups for Physicists} (Wiley, New York, 1974).

\bibitem{stoler70} D~Stoler, ``Equivalence Classes of Minimum Uncertainty Packets,'' Phys.\ Rev.\ D {\bf 1},  3217 (1970).

\bibitem{dollard-friedman77} J.~D.~Dollard and C.~N.~Friedman, ``Product integrals and the Schrödinger equation,'' J.\ Math.\ Phys.\ {\bf 18}, 1598 (1977).

\bibitem{dollard-friedman79} J.~D.~Dollard and C.~N.~Friedman, {\it Product Integral} (Addison-Wesley, Reading, MA, USA, 1979).

\bibitem{BCOR09} S.~Blanes, F.~Casas, J.~A.~Oteo and J.~Ros, ``The Magnus expansion and some of its applications,'' Phys.\ Rep. {\bf 470}, 151 (2009).

\bibitem{piazza-calucci02} A.~Di~Piazza and G.~Calucci, ``Pair production in a rotating strong magnetic field,'' Phys.\ Rev.\ D\ {\bf 65}, 125019 (2002).

\bibitem{kim-13b} S.~P.~Kim, ``Second Quantized Formulation of Scalar QED in Time-Dependent Magnetic Fields,'' in prepation.

\end{thebibliography}
\end{document}